\documentclass[aps,prb,showpacs,twocolumn,amssymb,amsmath,floatfix,superscriptaddress]{revtex4}
\usepackage{graphicx,bm}
\begin{document}
\bibliographystyle{prsty}

\title{Theoretical analysis of STM-derived lifetimes of excitations in the 
Shockley surface state band of Ag(111)}

\author{M. Becker}
\email{becker@physik.uni-kiel.de}
\affiliation{Institut f\"ur Experimentelle und Angewandte Physik, 
Christian-Albrechts-Universit\"at
zu Kiel, D-24098 Kiel, Germany}

\author{S. Crampin}
\email{s.crampin@bath.ac.uk}
\affiliation{Department of Physics, University of Bath, Bath BA2 7AY, 
United Kingdom}

\author{R. Berndt}
\affiliation{Institut f\"ur Experimentelle und Angewandte Physik, 
Christian-Albrechts-Universit\"at
zu Kiel, D-24098 Kiel, Germany}

\date{\today}

\begin{abstract}
We present a quantitative many-body analysis using 
the GW approximation
of the decay rate $\Gamma$ due to electron-electron scattering
of excitations in the Shockley surface state band of Ag(111),
as measured using the scanning tunnelling microscope (STM).
The calculations include the perturbing influence of the STM, which
causes a Stark-shift of 
the surface state energy $E$ and concomitant increase in $\Gamma$. 
We find $\Gamma$ varies more rapidly with $E$ than recently found
for image potential states, where the STM has been shown to significantly 
affect measured lifetimes. 
For the Shockley states, the Stark-shifts 
that occur under normal tunnelling conditions are relatively small
and previous STM-derived lifetimes need not be corrected.
\end{abstract}
\maketitle

The femtosecond lifetimes $\tau$ of electronic excitations in noble metal
surface bands \cite{ech04_} can be determined using the scanning tunnelling 
microscope (STM).  The STM-based techniques complement those that
use photo-excitation such as photoemission spectroscopy (PES) and 
two-photon photoemission \cite{ech04_},
being applicable to both electron-like and hole-like
excitations in combination with the advantages of atomic-scale
spatial resolution.
There are two principal methods available. The first is based upon
line shape analysis of differential conductivity ($dI/dV$) 
measurements \cite{jli98_,kli00a}, along with artificial or naturally 
occurring nanoscale resonator structures which are 
used to induce spectral structure 
over a wide range of energies \cite{kli01_,jen05_,cra05b}. 
The second approach is based upon the spatial variation
of quantum interference patterns that are visible in $dI/dV$ measurements
\cite{bur99_,bra02_,cra05a}.
These are analysed to determine the phase coherence length $L_\phi$, 
which can be converted to the lifetime via the group velocity $v_g$: 
$\tau=L_\phi/v_g$. 

An analysis of Ag(111) Shockley lifetime values 
determined using these 
two different STM techniques has demonstrated that considerable agreement 
exists, over a range of energies\cite{kro05_}. The lifetimes are also comparable to
state-of-the-art many-body calculations using the GW method, 
but decrease more rapidly with increasing energy.
In comparing theoretical and experimental lifetimes no consideration
has yet been given to the possible consequences of the perturbing
influence of the STM tip, which is known to cause a measurable 
Stark-shift in the surface state energies \cite{lim03_}.
Many-body GW calculations have shown that in the case of the
higher-lying image potential surface states this Stark-shift
is accompanied by a significant increase in the inelastic decay rate
\cite{cra05c}; the electric field between the STM tip and sample 
leads to a doubling of the decay rate under normal tunnelling conditions.
Here we perform a quantitative study using the many-body GW method 
to quantify for the first time the impact of the STM tip on the 
electron-electron scattering rate of Shockley state electrons.
We discount the possibility of a significant Stark-shift induced change to 
the electron-phonon contribution, which has previously been shown to be constant 
for excitations with energies in excess of 20 meV.\cite{eig03_}

Our calculations are based upon the approach developed by Chulkov and
coworkers \cite{chu98_}, and used widely in calculations of surface 
state dynamics\cite{ech04_} including the lifetimes of Stark-shifted 
image potential states.\cite{cra05c} 
The damping rate or inverse lifetime of an excitation in the
state $\psi(\bm{r})$ with energy $E$ is obtained from the 
expectation value of the imaginary part of the electron self-energy, 
$\Sigma(\bm{r},\bm{r}';E)$:
\begin{equation}
\Gamma=\tau^{-1}=
-2\int\!\! d\bm{r}\!\int \!\!d\bm{r}' \, \psi^{*}(\bm{r})\mathrm{Im}
\Sigma(\bm{r},\bm{r}';E)\psi(\bm{r}').
\label{eqn:gamma}
\end{equation}
In the GW approximation \cite{hed69_} of many-body theory the imaginary
part of the self energy is calculated in terms of the screened Coulomb
interaction $W$ and the Green function $G$
\begin{equation}
\mathrm{Im}\Sigma(\bm{r},\bm{r}';\epsilon)\!=\!
-\frac{1}{\pi}\int_{E_F}^\epsilon \!\!d\epsilon'\,
\mathrm{Im}G(\bm{r},{\bm r}';\epsilon')
\mathrm{Im}W(\bm{r},\bm{r}';\epsilon\!-\!\epsilon').
\label{eqn:imsig}
\end{equation}
A successful account of the decay rates of the noble metal
Shockley surface states is possible using for the Green function
its non-interacting counterpart
\begin{equation}
G(\bm{r},\bm{r}';\epsilon)=
\sum_{i}\frac{\psi_{i}(\bm{r})\psi_{i}^{*}(\bm{r}')}
{\epsilon-E_{i}+i0^+},
\label{eqn:g}
\end{equation}
and evaluating the screened Coulomb interaction in the random phase 
approximation (RPA),
\begin{eqnarray}
W(\bm{r},\bm{r}';\omega)&=&v(\bm{r}-\bm{r}')+\int d\bm{r}_1\int d\bm{r}_2
v(\bm{r}-\bm{r}_1)\nonumber\\
&&\times \chi^{0}(\bm{r}_1,\bm{r}_2;\omega)W(\bm{r}_2,\bm{r}';\omega)
\label{eqn:w}
\end{eqnarray}
where $v$ is the bare Coulomb interaction and $\chi^{0}$
is the density-density response function of the non-interacting electron
system. 

For the evaluation of the single particle states we have solved
the Schr\"odinger equation using standard
pseudopotentials varying in the direction perpendicular to the surface
\cite{chu99_,ef}.
The pseudopotential does not describe the $d$-electrons of the substrate,
which are anyway too low in energy to play a significant
role as final states for the decay,
but their contribution to the screening is included via
a polarisable background \cite{lie93_,lek03_}.

\begin{figure}[t!]
\includegraphics[bbllx=49,bblly=49,bburx=321,bbury=303,
width=85mm,clip=]{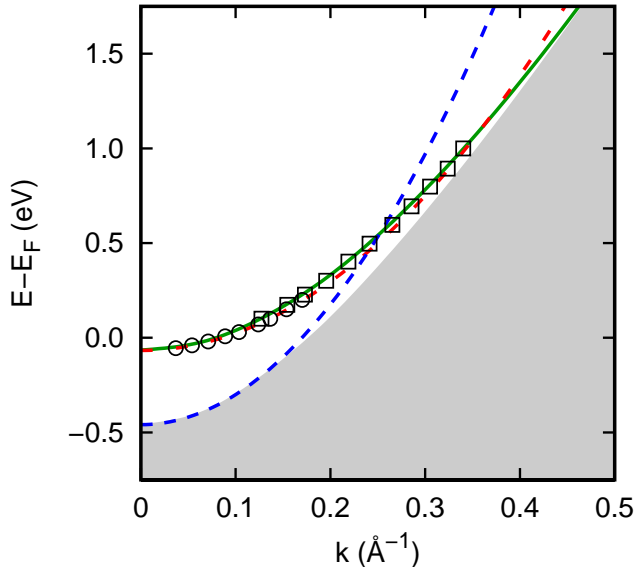}
\caption{
\label{fig:disp}
Surface band structure of Ag(111). The shaded area and solid line are the
projected continuum and surface state dispersion from {\it ab-initio} 
calculations which we use for calculating the lifetime.
Symbols are STM-derived dispersion data: circles, Ref.\ \onlinecite{jli97_};
squares, Ref.\ \onlinecite{vit03_}.
The dashed lines are parabolic dispersions with effective masses $m^\ast=0.42$
(surface state band) and 0.24 (band edge).}
\end{figure}

Previous calculations of the decay rate of the Ag(111) Shockley state
have assumed parabolic dispersion with effective masses of $m^\ast=0.42$
and $m^\ast=0.24$ for the intrinsic surface state and
the lower band edge respectively, but we find that over the extended energy
range of interest here these provide a poor 
description of the dispersion that is found in {\it ab-initio} calculations,
as may be seen in Fig. \ref{fig:disp}. The parabolic dispersions result in
the surface state crossing the band edge, which is incorrect.
In the calculations reported here we have therefore
used the {\it ab-initio} dispersions.
As previously noted \cite{vit03_,ver05_} there are important changes in the 
shape of the surface state wave function with $k$, the surface wave vector,
and we take these into account by recalculating the wave function
for different $k$ with the pseudopotential parameters changed to take into
account the appropriate {\it ab-initio} band edges and surface state energy.
Finally we have also modified the potential outside the surface in order 
to take account both of the 
multiple images present in the tunnel junction geometry and the 
electric field between tip and sample due to the applied bias voltage,
as described in Ref. [\onlinecite{lim03_}]. There it was found that
measured Stark-shifts were best described by assuming no contact potential
existed between tip and sample,
and so in the present calculations we have assumed that the tip and 
sample have similar work function.

We first consider the influence of the STM tip on the 
lifetimes of holes at the {\it bottom} of the Ag(111) surface state band 
($k=0$),
which have been determined experimentally from line shape analysis of the 
step-like onset that is seen in open-feedback $dI/dV$ measurements
on defect free regions of the surface.\cite{jli98_}
The onset occurs when the bias voltage coincides with the 
surface state binding energy, so that the Fermi energy of the tip coincides with the
minimum of the surface state band. This is Stark-shifted downwards from the 
value of $E_0=-63$ meV observed in PES experiments \cite{rei01_}
by the STM-derived electric field. The field itself varies according to the 
tip-sample separation, which depends experimentally upon the size of the current
prior to opening the feedback loop, so that different choices
for this current result in different shifts. Limot {\it et al.}\cite{lim03_}
were able to follow the change in $E_0$ from $-66$ to $-80$ meV using 
currents in the range 50 pA $-$ 6 $\mu$A.

We simulate these conditions as follows. For a given
tip-sample separation we solve the Schr\"odinger equation for the tunnel
junction to determine the energy of the surface state, varying the 
bias voltage until the calculated Stark-shifted binding energy and the
bias voltage coincide (threshold tunnelling condition).
At this point we use equations (\ref{eqn:gamma})--(\ref{eqn:w}) to calculate
the hole decay rate $\Gamma$ 
that applies for this value of the Stark shift, before 
repeating for different values of the tip-sample separation.

\begin{figure}[t!]
\includegraphics[bbllx=49,bblly=49,bburx=321,bbury=303,
width=85mm,clip=]{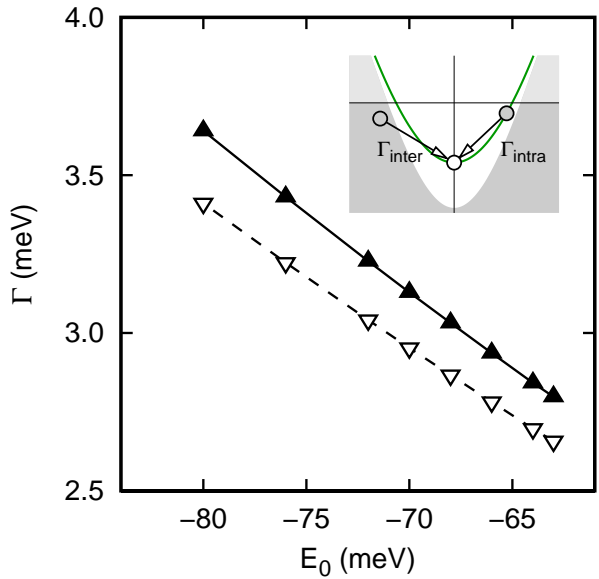}
\caption{
\label{fig:e0dat}
Decay rate $\Gamma$ due to electron-electron scattering 
of hole-like excitations at the bottom of the 
Shockley surface state band on Ag(111), as a function of Stark-shifted
energy $E_0$. Filled triangles show the total decay rate, and open triangles 
the intraband contribution. The inset illustrates intraband and interband decay 
processes.}
\end{figure}

The results are shown in Fig. \ref{fig:e0dat}.
Our field-free value agrees with the $2.8$ meV lifetime broadening
calculated by Garc\'{\i}a-Lekue {\it et al}.\cite{lek03_} using similar
methods. Including the field due to the STM tip,
we find a linear increase in the decay rate as the surface state is 
Stark-shifted to greater binding energies, with $\Gamma$ 
some 30\% greater when the binding energy is 
$E_0=-80$ meV than at the field-free energy of $-63$ meV. The 
rate of increase is comparable to that found for image state electrons,
with $|d\Gamma/dE_0|=0.050$, compared to $d\Gamma/dE_1=0.037$ for the Cu(001)
$n=1$ image state,\cite{cra05c} although the range of accessible
Stark-shifted energies is significantly smaller so that the absolute
change in decay rates that occurs for the Shockley state is much smaller.
The integral in Eqn. (\ref{eqn:imsig}) is over final states, and by
isolating contributions to the imaginary self energy due to the surface state 
band and using just those to calculate the decay rate allows $\Gamma$ 
to be decomposed into intraband and interband contributions.
We find that the variation in the decay rate shown in Fig. \ref{fig:e0dat} is
primarily due to the change in the rate of intraband scattering, 
the process by which the hole is filled by an electron from within the 
surface state band itself. The interband contribution, in which the hole is 
filled by an electron from the bulk continuum, 
contributes less than 6\% of the overall decay rate. 

To understand the origin of the change in the decay rate we have performed 
a number of auxiliary calculations, focusing on the dominant intraband decay 
rate. These show that the physical basis for the increase in the decay rate 
is the increased number of final states available for decay. 
The calculations are as follows. 
At the field-free surface we find $\Gamma_{\mathrm{intra}}=2.66$\, meV,
and when the surface state is Stark-shifted to a binding energy of $-80$ meV
we find $\Gamma_{\mathrm{intra}}=3.41$\, meV. If we calculate the decay rate
using the Stark-shifted eigenvalues but replace the corresponding
wave functions by their unperturbed counterparts, we find 
$\Gamma_{\mathrm{intra}}=3.37$\, meV, which is close to the result of
the calculation using the field-perturbed wave functions, and conversely
using the Shockley wave functions calculated in the presence of the 
applied field, but the unperturbed energies, gives 
$\Gamma_{\mathrm{intra}}=2.67$\, meV, close to the field-free decay rate.
Hence changes in the wave functions
of the surface and bulk states due to the STM-induced electric field
are of minor importance, 
which is different to the situation found for the
$n=1$ image state at Cu(001) where changes in the wave function are 
the most important contributor to the increased decay rate.\cite{cra05c} 

Next we have also calculated the decay rate using the field-free
energies and wave functions, but with the surface state wave function
in the Green function Eqn. (\ref{eqn:g}) scaled by $\sqrt{80/63}$; $80/63$ is 
the ratio of the number of electrons in the surface state band in the 
perturbed and unperturbed systems.
The resulting decay rate is $\Gamma_{\mathrm{intra}}=3.41$\, meV. This
simple re-normalisation of the contribution of the surface state to the
sum (integral) over final states in the evaluation of the self-energy,
Eqn. (\ref{eqn:imsig}), is sufficient to reproduce the result of the 
calculation using the properly Stark-shifted states,
and indicates that it is only the 
change in the total number of states available to fill the hole that
matters, and not their changed energies.

We have extended this study to also consider the decay rate of excitations at 
energies further up the Ag(111) Shockley surface state band.
These have been measured 
experimentally using line shape analysis in nanoscale 
resonators\cite{kli01_,jen05_}
and from studies of quantum interference patterns near steps\cite{bur99_}
and in triangular adatom corrals.\cite{bra02_}
\begin{figure}[t!]
\includegraphics[bbllx=49,bblly=49,bburx=321,bbury=303,
width=85mm,clip=]{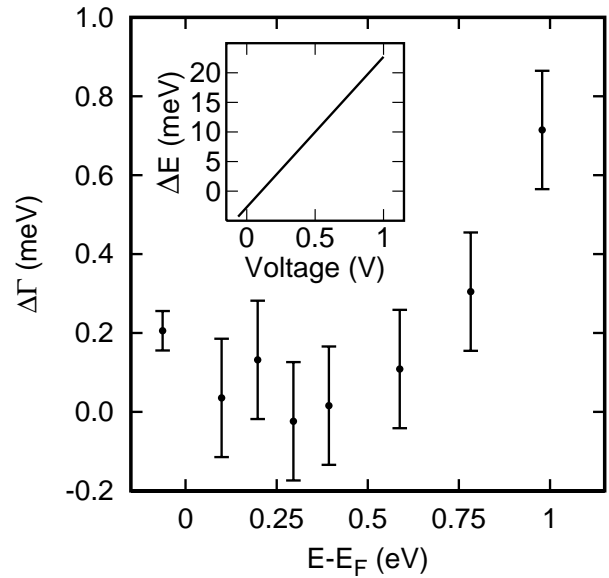}
\caption{
\label{fig:ekdat}
Calculated change in the decay rate induced by the electric field
of the STM, for Ag(111) surface state electrons at the tunnelling 
threshold ($E-E_F=eV$).
A tip-sample separation of 8\,\AA\, has been used.
The inset shows the size of the Stark-shift.}
\end{figure}
We consider energies between 0 and 1 eV where the surface state is well 
defined.  Physically this energy range is of interest as within it there is a 
cross-over in the primary decay channel from intraband to 
interband dominated decay.\cite{vit03_}
In these calculations we keep the tip-sample separation fixed at 8\,\AA\,
as the bias voltage is varied
(this distance results in a 4 meV Stark-shift of the band edge 
state, as was presented in Ref.\ [\onlinecite{kli00a}]), 
and we once again calculate the decay rate of the surface state
at the tunnelling threshold, where the energy of the Stark-shifted state 
coincides with the Fermi energy of the tip. 

The results are presented in Fig.\ \ref{fig:ekdat}.
Using this fixed tip-sample separation we find that the
Stark-shift $\Delta E$ of the surface state energy increases linearly with the 
bias over the range of voltages considered, and reaches 22 meV when the 
bias is 1 V (Fig.\ \ref{fig:ekdat}, inset)\cite{mi}. However, accompanying the
field-induced modification of the surface state we find a rather small
change in the decay rate. 
The additional numerical complexity associated with
the reduced symmetry of an excitation with a finite wave vector results in
a greater uncertainty in the calculated decay rate away from the band edge. 
Allowing for this, we find that the change in the decay rate is less than 
0.3 meV over most of the range, only reaching
$0.7\pm 0.15$ meV for a bias of 1 V.

We now consider our results as a whole and in the context of 
experimental studies.
In the case of holes at the bottom of the surface state band, 
in the measurements of Kliewer {\it et al}\, \cite{kli00a} the onset was observed
at $-67$ meV for Ag(111), corresponding to a Stark-shift of 4 meV from the PES 
field-free value.\cite{rei01_} 
Our calculations reported here (Fig. \ref{fig:e0dat}) show that the
the corresponding decay rate is increased by
0.2 meV above the field-free value, which is within the experimental 
uncertainty.\cite{cu} 
For electron-like excitations in the surface state band the decay rate
increases with energy due to the increase in available decay channels, 
and for an energy 1 eV above $E_F$ the corresponding contribution to the linewidth is approximately 40 meV.
The STM-induced increase of $\simeq 0.7$ meV is small in comparison, and again
well within the experimental uncertainty.\cite{vg}

Therefore, to conclude, on the basis of our quantitative theoretical many-body 
analysis using the GW approximation we have demonstrated 
that in contrast to the case of image 
potential states, the lifetimes of excitations in noble metal Shockley 
surface state bands determined using the STM are not sufficiently affected 
by the electric field of the STM tip that previously reported values need
to be corrected.
We have identified that relative changes of up to 30\% can occur
under extreme but experimentally accessible tunnelling conditions,
and the
decay rate actually 
changes more rapidly as a function of the Stark-shift for Shockley states
than for image potential states. 
However the significantly smaller Stark-shifts 
that occur when tunnelling via the lower-lying Shockley states 
under normal tunnelling conditions
mean that the absolute changes in the decay rate lie within
experimental uncertainties previously reported. 

\begin{acknowledgments}
M.B. and R.B. thank the Deutsche Forschungsgemeinschaft for financial support, and E. Pehlke for discussions.
\end{acknowledgments}


\begin{thebibliography}{10}
\bibitem{ef}
For Cu(111) and Ag(111) the Fermi energies from 
Ref. \onlinecite{chu99_} were modified
to 7.014 and 5.085\, eV respectively so as to reproduce the 
most recent field free (PES) values for the energies of the 
Shockley states of $E-E_F=-435$ and $-63$ meV.
\bibitem{mi}
The Stark-shift towards greater binding energies for vanishing voltages is due to the multiple images present in the STM-geometry.
\bibitem{cu}
We have performed a similar set of calculations to those reported in detail 
here for the corresponding Shockley state at Cu(111),
where the field-free binding energy is significantly greater at
$-435$ meV [\onlinecite{rei01_,kro04_}].
Tunnelling spectroscopy of the band edge state therefore requires larger
bias voltages which result in 
a relatively greater Stark-shift then at Ag(111). For Cu(111) we also find 
$|d\Gamma/dE_0|=0.050$, so that under typical
tunnelling conditions [\onlinecite{kli00a,kro04_}] the field-induced decay rate 
increases by less than 0.5 meV, a small fraction of the full line width
of 24 meV and within the experimental uncertainty.
\end{thebibliography}
\end{document}